\documentclass[prd,a4paper,aps,epsfig,showpacs,twocolumn,floats]{revtex4}

\newcommand{\be}{\begin{equation}}
\newcommand{\ee}{\end{equation}}
\newcommand{\bea}{\begin{eqnarray}}
\newcommand{\eea}{\end{eqnarray}}

\newcommand{\db}[1]{\overline{\delta{#1}}}
\newcommand{\Ha}{\mathcal{H}}
\newcommand{\Ga}{\mathcal{G}}
\newcommand{\ep}{\varepsilon}
\newcommand{\pr}{^{'}}
\newcommand{\dbu}{\db{u}_{\parallel}}
\newcommand{\ct}{{c_s}^2}

\newcommand{\ba}{\begin{align}}
\newcommand{\ea}{ \end{align} }
\newcommand{\bs}{\begin{split}}
\newcommand{\es}{\end{split}}

\begin{document}
\title{Primordial fluctuations without scalar fields}
\author{Jo\~ao Magueijo and Johannes Noller }
\affiliation{
Theoretical Physics Group, Imperial College, London, SW7 2BZ}
\date{\today}
\begin{abstract}
{We revisit the question of whether fluctuations in
hydrodynamical, adiabatical matter 
could explain the observed structures in our 
Universe. We consider matter with variable equation of state $w=p_0/\ep_0$
and a concomitant (under the adiabatic assumption) density dependent
speed of sound, $c_s$. We find a limited range of possibilities for a set
up when modes start inside the Hubble radius, then leaving it and freezing 
out. For expanding Universes, power-law $w(\ep_0)$ models
are ruled out (except when $c_s^2\propto w \ll 1$, requiring
post-stretching the seeded fluctuations); but sharper profiles in $c_s$
do solve the horizon problem. Among these, a phase transition in $c_s$ 
is notable for leading to scale-invariant fluctuations if the initial 
conditions are thermal. For contracting Universes all power-law $w(\ep_0)$
solve the horizon problem, but only one leads to scale-invariance:
$w\propto \ep_0^2$ and $c_s\propto \ep_0$. This model 
bypasses a number of problems with single scalar field
cyclic models (for which $w$ is large but constant).}
\end{abstract}
\pacs{0000000}
\maketitle

\bigskip

\section{Introduction}
Scalar fields are an easy tool for modeling the early universe, 
justifying their popularity. Unfortunately we have yet to detect
a fundamental scalar field, and should one be found in high energy
physics experiments this hardly proves their case in cosmology. 
In this paper we consider models for primordial structure formation 
based on hydrodynamical matter subject to certain thermodynamical constraints. 
Our construction is not dissimilar to that of~\cite{hu,jochen}, where the 
concept of ``generalized dark matter''---a parametrization of a generic
hydrodynamical fluid---was introduced to explain the current cosmic 
acceleration. Our motivation, however, concerns 
the generation of a (quasi) scale-invariant spectrum of density 
fluctuations in the early universe.

Generating primordial density fluctuations entails a solution to the
horizon problem, since the scales we now observe are initially 
causally disconnected according to the unreformed Big Bang model.  
A number of solutions have been proposed. Inflation is often 
invoked~\cite{infl}, and vacuum quantum fluctuations inevitably present 
in the modes inside the Hubble radius are transformed into
near-scale invariant fluctuations near the deSitter phase.  
Other possibilities include the cyclic or ekpyrotic 
scenarios~\cite{ekp}, string gas cosmology~\cite{hag} and the varying 
speed of light (or sound) framework~\cite{csdot}. Combinations of these
possibilities have also been considered (see for example~\cite{piazza}).

If we are to replace the proverbial scalar field with a hydrodynamical 
fluid the obvious question is what is its equation of state?
In this paper we consider a power-law dependence between $w$ and
density $\rho$. This is a feature of the popular Chaplygin gas
and its modifications~\cite{chap}, but also
of intermediate inflationary  models~\cite{barr,and}. 
In fact the latter leads to a second solution for scale-invariance 
in the inflationary setting (the other solution 
being slow-roll inflation~\cite{staro}).

It is important to stress that hydrodynamical fluids 
and scalar fields are intrinsically different. Stress-energy fluctuations 
on scalar fields result from a cross term between zero and first order
field fluctuations. The zeroth component is never thermalized even if the
fluctuations are. In contrast, a thermal fluid behaves as a single
unit, and so a number of thermodynamic theorems valid for fluids don't apply
to scalar fields. This should be borne in mind throughout this paper 
(see~\cite{pedro} for an illustration of this point). The formalism
of cosmological perturbations with a varying $w$ has been studied before
but only in the context of radiation and matter mixtures (and multifluids
more generally~\cite{kodama,mukh,lidsey}.)

The plan of our paper is as follows. In Section~\ref{background} we review
results for the background solution with varying $w$ needed for the
rest of this paper. Then in Section~\ref{horgauge} we examine the 
hypothesis of adiabaticity and spell out its implications for the horizon
and gauge problems. The formalism of cosmological perturbation theory
for adiabatic fluctuations with a varying $w$ is developed in 
Section~\ref{perteqs}. A number of solutions are presented and
discussed in Section~\ref{sls1} and \ref{sls2}. 
A summary of our results is given
in a concluding section. 

Throughout this paper we shall assume that the fluid satisfies the 
adiabatic conditions. Non-adiabatic fluctuations are the subject of
a paper in preparation~\cite{nonad}.

\section{Background solutions with varying $w$}\label{background}
To set the notation let the Friedmann equations be written as:
\be
\Ha^2 = \frac{1}{3} a^2 \ep_{0}  
\ee
\be
\Ha^2 - \Ha\pr = \frac{1}{2} a^2 (\ep_0 + p_0) \equiv \Ga, 
\ee
where $\Ha=a'/a$ (with $a$ the expansion factor and 
a prime denoting derivative with respect to conformal
time) and where
we have defined the variable $\Ga$ 
in order to keep notation concise in what follows.
The Friedmann equations imply a conservation equation for $\ep$
\be
\ep_0\pr = - 3 \Ha (\ep_0 + p_0)\; .
\ee 
These equations establish the background dynamics in units where
$8\pi G=1$. In order to solve them an equation of state 
$w=p_0/\ep_0$ must be specified.

A number of background solutions with varying $w$ can be found in 
the literature. In this paper we shall focus on an adaptation of
the equation of state~\cite{barr}:
\be\label{w1}
1+w=\left(\frac{\ep_0}{\rho_\star}\right)^{2\beta}\; ,
\ee
with $\beta>0$. This leads to exact solution:
\bea
a&=&a_0\exp{(t/t_\star)^q}\, ,\label{aeq}\\
\ep_0&=&\frac{3q^2}{t_\star^2}{\left(\frac{t}{t_\star}\right)}^{2q-2}
\label{rhoeq}
\eea
with parameters $q$ and $t_\star$ related to model parameters 
$\rho_\star$ and $\beta$ via:
\bea
q&=&\frac{4\beta}{1+4\beta}\\
\rho_\star &=& \frac{3q^2}{t_\star^2}(2\beta)^{\frac{1}{2\beta}}
\sim \frac{1}{t_\star^2}
\; .
\eea
The above solutions have been given in terms of proper time,
rather than conformal time.
The above excludes the case $\beta=0$, and $\beta=-1/4$,  $q=-1/2$, 
that is $w\propto1/\sqrt{\ep_0}$; but this case is outside
our $\beta>0$ assumption anyway.

The regime $t\gg t_\star$ and $\ep_0\ll\rho_\star$ corresponds to 
intermediate inflation, one of the two inflationary solutions
for scale-invariant fluctuations~\cite{barr,and,staro}.
We are interested in $w\gg 1$ (associated with 
$c_s\gg 1$ in hydrodynamical fluids) and thus we'll focus on the opposite 
regime: $\ep_0\gg \rho_\star$ and $t\ll t_\star$. More generally
we can consider a variety of other models, with:
\be\label{w2}
w=w_0 + \left(\frac{\ep_0}{\rho_\star}\right)^{2\beta}
\ee
where $w_0$ is the fluid's low energy equation of state. At high
density these models all have the same (power-law) behaviour.
The high $w$ phase then exits into a constant $w$, low density phase
which need not be inflation. 
We can also consider contracting models with these equations
of state: an obvious generalization of cyclic
models, where a high, but constant $w$ is invoked.

In the phase with $\ep_0\gg \rho_\star$ and $w\gg 1$ we have that:
\bea
a&\approx& a_0\\
\ep_0&\propto& t^{2q-2}.
\eea
This highlights the main peculiarity of these models. 
As a function of time the universe appears to be loitering 
($a\approx a_0$) and $a$ doesn't vanish at the Big Bang
(although its time derivative diverges).
However, its density changes like a power-law in $t$ and diverges
as $t\rightarrow 0$. So although the scale factor doesn't vanish
at $t=0$ we do have a Big Bang singularity.
The fact that the metric and the matter change on different
time-scales is behind a number of subtleties to be examined in
this paper. Notice, for example, that conformal and proper time
may be used interchangeably in many calculations.

\section{Adiabatic fluctuations and the horizon and gauge
problems}\label{horgauge}
For the purpose of this paper a ``solution of the horizon problem'' means
the existence of a phase in the early universe when modes are initially
inside the Hubble radius (ruled by causal micro-physics), then cross
outside to become dominated by gravity (i.e. ``freezing-out'').
This is the opposite kinematics to that realized in the
standard Big Bang model. A number of solutions have been proposed, 
most notably inflation~\cite{infl}, ekpyrotic models~\cite{ekp}, 
a Hagedorn phase~\cite{hag} and a varying speed
of light/sound~\cite{csdot}. Constant and varying $w$ models should be discussed
separately if we examine the issue for adiabatic fluids.

Should $w$ and $c_s$ be constant, adiabaticity 
implies $w=c_s^2$, as is well known. This implies $w>0$, 
precluding inflation. Inflation can never be realized by a single
adiabatic fluid, and it's not an accident that scalar fields are
preferred. Thus, for adiabatic fluids, the horizon problem has
to be solved with a contracting phase (e.g. in cyclic models~\cite{ekp}) or
with a phase transition in the early Universe after a loitering
epoch~\cite{hag}. With a varying speed of sound, the adiabatic
assumption requires a varying $w$ (in contrast with~\cite{csdot})
which we proceed to discuss.

If $w$ is variable, adiabaticity imposes a more complex relation than
$w=c_s^2$. One can study the perturbation equations for modes 
inside the horizon (ignoring expansion) and derive the speed of
propagation of pressure waves from first principles. This should
be the rightful definition of
the speed of sound $c_s$ and can be computed to be:
\be\label{csdef}
c_s^2 =\frac{\delta p}{\delta \ep}\; ,
\ee 
where $\delta p$ and $\delta \ep$ are the perturbed pressure and energy
density.
If this derivative is adiabatic, and if the background evolution is 
also adiabatic, then this implies the differential equation between
$w$ and $c_s$:
\be\label{adiab}
c_s^2=\frac{d(w\ep_0)}{d\ep_0}=w+\ep_0 w_{,\ep_0}\; .
\ee
More generally (\ref{adiab}) is true whenever the same 
equation of state $p=p(\ep)$ is valid for fluctuations and unperturbed
fluid (this isn't true, e.g. for scalar fields~\cite{pedro}). Note,
however, that in non-adiabatic situations the correct definition
for the speed of sound is (\ref{csdef}) and {\it not} the definition given
in terms of background quantities (e.g.~\cite{bardeen}), which leads
to (\ref{adiab}).

Assuming that (\ref{adiab}) is verified, if $w$ is a power law 
in $\ep_0$ then we still have $c_s^2\propto w$ (even though $c_s^2\neq w$).
The variation in $w$ can 
be ignored only if $c_s$ is a step function (or is ruled by a very large power).
Thus the importance of investigating perturbations in 
varying $w$ fluids if one appeals to a varying $c_s$ to solve the
horizon problem.
 More generally we can consider non-adiabatic phenomenological laws:
\bea
c_s&\propto& \ep_0^\alpha\\
w&\propto&\ep_0^{2\beta}
\eea
with the adiabatic case resulting from $\alpha=\beta$ and a
suitable adjustment of the proportionality constants. Models
with $\alpha\neq \beta$ will be the subject of a separate paper~\cite{nonad}.

At this point we wish to highlight a subtlety, capable of explaining
many of the results presented later in this paper. 
The gauge-problem in varying $w$ models becomes distinct from 
the horizon problem and imposes constraints on 
$\alpha$ and $\beta$. Usually sub-horizon
modes do not suffer from the gauge-problem, whereas super-horizon modes do.
Thus if the horizon problem has been solved, 
setting up initial conditions is gauge-invariant. 
By this we mean that it only involves
scales for which gauge-transformations have negligible effect, 
so that different choices 
of gauge fixing (or different ``gauge-invariant'' options for $\delta\ep$,
for example), are equivalent (see~\cite{stew} and specially~\cite{durrer}).

But in varying-$w$ models this is no longer true. 
Gauge transformations are generated
by the Lie derivative of the background quantities~\cite{kodama,stew}.
The background metric and matter variables now change
on different time scales (cf. discussion at the end of 
Section~\ref{background}), specifically:
\be
\frac{\ep_0'}{\ep_0}\sim w\frac{a'}{a}
\ee
and we highlight the extra $w$ factor in the right hand side.
Therefore gauge transformations have negligible effect on different
scales for metric and matter variables. 
For metric variables gauge transformations
are irrelevant for modes satisfying:
\be
c_s k\gg \frac{a'}{a}
\ee
whereas for energy and pressure fluctuations the criterion is
\be
c_s k\gg \frac {a'}{a} w\; .
\ee
If we don't want the initial conditions to be plagued by a gauge problem 
we now have a more stringent condition than the criterion for solving the
horizon problem. For metric variables this reduces to $\alpha>1/2$,
which is indeed the condition for solving the horizon problem.
But for matter variables (in the models under consideration, described
in Section~\ref{background}) we have
\be
\alpha>\frac{1}{2}+2\beta
\ee
(we have used approximations, such as $a\approx$const, peculiar to
these models). This condition can never be met by adiabatic models.

The above assumes that the universe is expanding. If it is contracting
the situation is reversed, as the universe gets denser and hotter
in time rather than diluting and cooling. So it would appear that
only contracting adiabatic models with varying $w$ can be used
for structure formation, at least withing the models considered. 
This will indeed be our conclusion, after detailed calculations.

\section{The linearized perturbation equations}\label{perteqs}
We now develop the machinery of cosmological perturbation theory
with a varying $w$. Past literature has been restricted to 
matter-radiation and other fluid mixtures~\cite{kodama,mukh,lidsey}. The
generalization to any varying $w$ is straightforward but non-trivial, 
particularly if the adiabatic assumption is dropped. 

The linearised perturbed Einstein equations for a perfect fluid 
stress-energy tensor of the form
\be
{T^{\alpha}}_{\beta} = (\ep + p)u^{\alpha}u_{\beta} - p{\delta^{\alpha}}_{\beta}
\ee
are
\be
\Delta \Phi - 3\Ha(\Phi\pr + \Ha\Phi) = \frac{1}{2} a^2 \db{\ep},
\label{scalar1}
\ee
\be
{{(a\Phi)}_{,i}}\pr = \frac{1}{2} a^2 (\ep_{0} + p_{0}) {\dbu}_i = \Ga {\dbu}_i,
\label{scalar2}  
\ee
\be
\Phi^{''} + 3\Ha \Phi\pr + (2\Ha\pr + \Ha^{2})\Phi = \frac{1}{2} a^2 \db{p}.
\label{scalar3}
\ee
where $\Phi$ is the Newtonian potential and the stress-energy fluctuations are
evaluated in the longitudinal gauge (denoted by an overbar); we refer the
reader to~\cite{mukh} for notation and further explanations. 
We use equation (\ref{scalar2}) to solve for $\Phi^{'}$ and $\Phi^{''}$:
\be
\Phi\pr = \frac{\Ga}{a} \dbu - \Ha \Phi \label{phi'}
\ee
\be
\Phi^{''} = \frac{\Ga}{a}\dbu\pr + \frac{\Ga}{a}{\left(\frac{p_0\pr}{\ep_0 + p_0} - 3 \Ha\right)} \dbu + \Ga \Phi  \label{phi''}
\ee
Combining equations (\ref{scalar1}) and (\ref{scalar3}), and
considering that ${c_s}^2 = \frac{\db{p}}{\db{\varepsilon}}$ 
we get the $\Phi$ equation:
\begin{equation}
\Phi^{''} +3(1 + {c_s}^2) \Ha \Phi\pr + (2\Ha\pr + (1 + 3{c_s}^2)\Ha^2 - {c_s}^2 \Delta)\Phi = 0\; .\label{Phieq}
\end{equation}
Using (\ref{phi'}) and (\ref{phi''}) to substitute in 
for $\Phi^{'}$ and $\Phi^{''}$ we also derive the
``equation of motion'' for $\dbu$
\begin{equation}
\dbu\pr + \dbu {\left(\frac{p\pr_0}{(\ep_0 + p_0)} + 3 {c_s}^2 \mathcal{H}
\right)} = {\left(1 + \frac{{c_s}^2\Delta}{\Ga}\right)}(a\Phi) 
\; . \label{ueq} 
\end{equation}
Equations (\ref{Phieq}) and (\ref{ueq}) are the precursors for the 
equations for variables ``$u$'' and ``$v$'' favoured in 
the literature~\cite{mukh,lidsey}. For fluids $\Phi$ is related to ``$u$'' and
the velocity perturbation $\dbu$ to ``$v$'' (or the related curvature
perturbation $\zeta$). These variables
will be used in a future publication~\cite{nonad}
to derive the second order action for the fluctuations with and without the
adiabatic assumption. Here however we take another route.

Should we assume adiabatic fluctuations a number of simplifications 
are possible. Since for adiabatic fluctuations 
${c_s}^2 = \frac{p_0\pr}{\ep_0\pr}$ we can derive the valuable
identity:
\be
\frac{p_0\pr}{\ep_0 + p_0} + 3 \ct \Ha = \frac{w'}{1+w}
+3(c_s^2-w)\Ha=0\; .
\ee
Eqn.~(\ref{Phieq}) can then be rewritten as an equation which 
for modes outside the horizon represents a conservation law. The 
``conserved quantity'' is 
\be
\zeta=\Phi\frac{5+3w}{3(1+w)}+\frac{2}{3(1+w)}\frac{\Phi'}{\Ha}
\label{zetadef}
\ee
and it's straightforward to prove that
\be\label{zetaprime}
\zeta'=-\frac{2}{3}\frac{c_s^2k^2\Phi}{\Ha(1+w)}
\ee
is equivalent to (\ref{Phieq}) for {\it any} functional dependence 
of $w$. When the 
pressure is negligible we obtain a constant $\zeta$, as announced.
The potential $\Phi$ can be eliminated altogether from this equation
by taking another derivative. After a few algebraic manipulations, 
we get:
\be
\zeta''+2\frac{z'}{z}\zeta' +c_s^2 k^2 \zeta=0\label{zetaeq}
\ee
with
\be\label{zeq}
z\propto \frac{a(1+w)^{1/2}}{c_s}\; .
\ee
Defining variable $v$ by $\zeta=v/z$, 
this equation becomes the more familiar:
\be\label{veq}
v''+{\left(c_s^2k^2-\frac{z''}{z}\right)}v=0\; .
\ee
The quantity $v$ 
is the variable which is ruled by a scalar
field action, when the second order action is evaluated. 

Equations 
(\ref{zetaeq}) and (\ref{veq}) reflect the tension between the 
dynamics for choices
$\zeta$ and $v$: For the first we have a friction term but no mass; 
for the latter no friction term and a (time-dependent) mass term. 
Sometimes it's easier to use one, sometimes the other. We shall keep
both on the plate, and rather than examine a ``$u$'' equation, infer all
the information about the potential $\Phi$ from $\zeta$ and equations
(\ref{zetadef}) or (\ref{zetaprime}).

\section{Expanding solutions}\label{sls1}
We now consider solutions to these equations separating the cases of
expanding and contracting Universes. In expanding Universes we find a 
no-go for all power-law models {\it with large} $w$
described in Section~\ref{background}.
There are, however, other models which bypass this negative result,
and display scale-invariant fluctuations. One example is
a Universe with $c_s^2=w\ll 1$ (but this requires follow-up inflation
or some other mechanism for stretching the size of the fluctuations).
Another successful adiabatic model is a 
phase transition in $c_s$, provided the initial fluctuations are thermal.

\subsection{Absence of solutions for power-law $w$}\label{sls1a}
Were it not for the discussion on the gauge-dependence of pressure waves 
in Section~\ref{horgauge} the following result might come as a surprise. 
We find that for the purpose of structure formation 
the horizon problem cannot be solved for {\it all} expanding 
{\it adiabatic} models with power-law $w(\ep_0)$ and $w\gg1$. 

This can be easily seen from the analysis of the two competing
terms in Eq.~(\ref{veq}). If $w\gg 1$, then  Eqn.~(\ref{zeq})
implies that $z\propto a$, so that the mass term is: 
\be
\frac{z''}{z}= \frac{a''}{a}=\frac{1}{6} a^2(\ep_0-3p_0)
\approx -\frac{1}{2}a^2w\ep_0.
\ee 
But the adiabatic condition (\ref{adiab}) applied to power-laws
implies that $c_s^2\propto w$, as we've seen. Thus, for any 
pure power-law, the effect of a varying
$c_s$ in the pressure term $c_s^2 k^2$ is exactly canceled
by the effect of $w$ on the varying mass term. Horizon
crossing occurs at $k_h^2\sim a^2 \ep_0$ and the strong energy condition
must be violated for this to increase in time. Specifically:
\be
\frac {(a^2\ep_0)^.}{a^2\ep_0}=-\frac{\dot a}{a}(1+3w)>0
\ee
so that $w<-1/3$ is needed.

\subsection{Near-dust with varying speed of sound}
It is possible to bypass this negative result in a number of cases,
but the extra requirement of scale-invariance narrows down two
possibilities. 

One assumption that may be dropped is $w\gg 1$ (as was first 
pointed out in~\cite{amend}; see also~\cite{piazza}). Specifically
we could consider model (\ref{w2}) with $w_0\approx 0$ and $\beta>-1/2$
in the regime $\ep_0\ll \rho_\star$. Then $z\propto a/c_s$ 
(i.e. $c_s$ and $w$ no longer cancel in $z$, as with $w\gg 1$). 
In addition $w$ is approximately constant ($w\approx 0$) and so 
we fall into the regime already studied in~\cite{csdot}. It was
shown that a solution for scale invariance is $c_s\propto\ep_0$,
so $\beta=1$ and $w=c_s^2/5\propto\ep_0^2$. 

While this represents a solution to scale-invariant 
structure formation the scales produced are too small. This can
be remedied by a mechanism stretching them following
the varying $w$ phase that seeded the fluctuations. 
It was suggested in~\cite{amend} that 
inflation might play such a role. But other methods of renormalizing 
the length scales could be attached to this model, such as a sharp
drop in the speed of light~\cite{am}, or a phase transition in the quantum
structure of space-time~\cite{holo}.

\subsection{A phase transition in $c_s$}
If one drops the power-law assumption in (\ref{w2}) altogether, 
the argument against
solving the horizon problem in expanding universes for adiabatic fluids
breaks down. A number of solutions satisfying
\be
{\left(\frac{c_s^2 z}{z''}\right)}'<0
\ee
are then possible. In general (\ref{adiab}) does not imply $c_s^2
\propto w$: this is a peculiarity of power-laws.   For example if
$c_s$ undergoes a phase transition:
\bea
\; {\rm   for} \; \ep_0<\rho_\star: \; c_s&=&{c_{s_-}}\\
\; {\rm   for} \; \ep_0>\rho_\star: \; c_s&=&{c_{s_+}} 
\eea
with $c_{s_+}\gg c_{s_-}$, then 
the adiabatic condition (\ref{adiab}) implies instead:
\bea
\; {\rm   for} \; \ep_0<\rho_\star: \; w&=&{c^2_{s_-}}\\
\; {\rm   for} \; \ep_0>\rho_\star: \; w&=&{c^2_{s_+}}+
(c^2_{s_-}-{c^2_{s_+}}) \frac{\rho_\star}{\ep_0}\; .
\eea
Therefore $w$ can be assumed to be a constant while $c_s$ is sharply
varying and Eqn.~(\ref{zeq}) implies $z\propto a/c_s$ 
(i.e. $c_s$ and $w$ don't cancel out in $z$). 
For $\ep_0\gg \rho_\star$, the model satisfies
(\ref{w2}) with $\beta=-1/2$; however, for $\ep_0\approx \rho_\star$ 
(relevant when $c_s$ is varying and structures are being seeded) we
can treat $w$ as a constant (as indeed $a$ and $\rho$ can be treated
as constants). We then fall into the case already studied in~\cite{csdot}.
The horizon problem is solved: modes start inside the horizon. If
their initial conditions are set by the zero-point vacuum fluctuations we
obtain a blue spectrum, with $n_s=2$; but if thermal initial conditions
are used,  we get scale invariance~\cite{csdot}.

There are other non-power law solutions for $w$ and $c_s$ that solve
the horizon problem for adiabatic fluids: for them the effect 
of $c_s$ in the pressure term no longer cancels with that of $w$ in 
$z''/z$. In general Eqn.~(\ref{adiab}) 
has a homogeneous solution, i.e. a $w$ profile which has no effect on $c_s$.
This is a Chaplygin-type of gas, with $w=A/\ep_0$. It is 
precisely this solution that appears linking the
two constant $c_s$ phases in the phase transition
scenario.

\section{A solution for cyclic scenarios}\label{sls2}

The situation described in the previous section is reversed should 
the universe be contracting, such as in cyclic models. 
Then for all power-law models described in Section~\ref{background}
modes start inside the horizon, then leaving and freezing
out. It's not difficult to work out the condition for approximate
scale-invariance. 

\subsection{The set up for the calculation}
A calculation similar to that in~\cite{csdot} can be applied
to these models. However, the usual Bessel solutions, interpolating 
between inside and outside the horizon regimes, are no longer valid.
Fortunately matching these regimes is enough for obtaining the spectrum and
amplitude of fluctuations left outside the horizon.

While the pressure term dominates the solution has the WKB form:
\be\label{solin}
\label{bc1} v\sim\frac{e^{ik\int c_s d\eta}}{\sqrt{2 c_s k}}\; ,
\ee
which acts as a boundary condition. The normalization ensures that, upon
second quantization, amplitudes  multiplying this solution become creation
and annihilation operators. As the Universe contracts and $\ep_0$ increases, 
the modes become dominated by the term in $z''/z$  in equation (\ref{veq})
(c.f. the argument given in Section~\ref{sls1a}). 
In this regime the general solution for (\ref{zetaeq})
(or for (\ref{veq})) is:
\be\label{solout}
\zeta=\frac{v}{a}=A +B\int\frac{d\eta}{a^2}\approx A+\frac{B}{a_0^2}
|\eta|,
\ee
where $A$ and $B$ are (possibly $k$ dependent) constants, and we
have used the  discussion in Section~\ref{background}
(cf. Eqn.~(\ref{aeq})) in the last approximation (note also that
$z\propto a$ approximately).

Usually the growing and decaying solutions are reversed in a 
contracting universe, and the constant, frozen-in mode becomes 
sub-dominant, while the time dependent mode diverges. 
This doesn't happen here: the term in $B$ actually goes to zero
as we approach the crunch. Throughout $\zeta\sim v$ since $a$
changes by at most by a factor of 2 throughout the relevant phase.

\subsection{Conditions for (near) scale-invariance}
The power spectrum left outside the horizon (fixed by factor $A$)
can be found  by gluing (\ref{solin}) and (\ref{solout}) at horizon crossing. 
Horizon crossing occurs when the two terms in $v$ in (\ref{veq})
are of the same order, i.e. when $c_s^2k^2=|z''/z|$. Using approximations
valid when $w\gg 1$ we have $k_h^2\propto a^2 w\ep_0/c_s^2$ so that
$k_h\propto \sqrt\ep_0$
for the gluing point. We take the ansatz
\be
\zeta (k,\eta)\approx v(k,\eta)\propto k^{\frac{n_s}{2}-2}\; 
\ee
for the constant $v$ left outside the horizon (in general this need
not be a power-law in $k$).  This should be glued to:
\be
v\sim\frac{1}{\sqrt{c_s k}}
\ee
when $k=k_h\propto\sqrt \ep_0$. Writing the resulting identity in 
terms of $\ep_0$ therefore produces the relation 
$\alpha+n_s/2-3/2=0$.
Gluing therefore implies the 
expression for the spectral index:
\be
n_s=3-2\alpha
\ee
(up to corrections logarithmic in $k$). The condition for scale-invariance
is $\alpha=1$, that is $c_s\propto \ep_0$ (just like
in models with constant $w$ discussed in~\cite{csdot}). 
The model associated with scale-invariance is therefore characterized by
the high density equation of state:
\be
c_s^2=3w = 3{\left(\frac{\ep_0}{\rho_\star}\right)}^2\; .
\ee
It  has expansion factor and density profile of the form:
\bea
a&=&a_0\exp{(|t|/t_\star)^{4/5}}\; ,\\
\ep_0&=&\frac{3(4/5)^2}{t_\star^2}{\left(\frac{t_\star}{|t|}\right)}^{2/5}\; .
\eea
The amplitude of the fluctuations is tuned by  parameter
$\rho_\star/M^4_{pl}$.

The above assumes a vacuum expectation value, producing a constant, 
$k$-independent factor when creation and annihilation operators
are inserted in second quantized solutions. If instead a thermal state is
taken as a boundary condition, then a factor of $T_c/k$ multiplies 
the spectrum, with $T_c=Ta$ the conformal temperature
 (cf.~\cite{pedro,csdot}). Reworking the result therefore
leads to a spectrum with index $n_s=2(1-\beta)$. Scale invariance under
thermal fluctuations thus requires  $\beta=1/2$ (which is the model
leading to intermediate inflation, but in a totally different regime).
We have $c_s\propto \sqrt{\ep_0}$ and $c_s^2=w$ and the model is
characterized by: 
\bea
a&=&a_0\exp{(t/t_\star)^{3/2}}\; ,\\
\ep_0&=&\frac{3(4/5)^2}{t_\star^2}{\left(\frac{t_\star}{t}\right)}^{2/3}\; .
\eea
Again the amplitude of the fluctuations is tuned by 
$\rho_\star/M^4_{pl}$, as well as $T_c$.

\subsection{Advantages over more standard cyclic scenarios}
The above has obvious similarities with traditional cyclic 
models~\cite{ekpperts}. In these $w\gg 1$ 
and $c_s=1$; here both $w$ and $c_s$ are large and vary like power-laws
of the energy density. 
Cyclic models, in their original formulation, have a number of 
possible problems~\cite{ekpperts}. It's the potential $\Phi$ that's
scale-invariant, not 
$\zeta$, which has a blue spectrum, with $n_s=3$. In addition $\Phi$
is not constant, but diverges like $1/|\eta|$. 
Recent models, with multiple scalar fields have attempted to fix these
shortcomings~\cite{multiekp}.

In desirable contrast, in the model we have 
proposed $\zeta$ is constant and scale-invariant. 
However the spectrum of $\Phi$ is not scale-invariant.
In order to obtain $\Phi$ we use (\ref{zetaprime}) for which we need to
know the next order expansion in $\zeta$. This is:
\be\label{zeta2norder}
\zeta=A k^{-3/2}(1+C_1(k\eta)^2 +C_2 k\eta)\; .
\ee
The first term comes from the first order corrections induced by a WKB
inclusion of the pressure terms; the second is the decaying mode calculated
for our model in (\ref{solout}). 
For any model the Taylor expansion in $k\eta$ will have two such leading
terms. If the first term dominates, $\zeta$ and
$\Phi$ have the same spectrum (as in inflation and the varying speed of sound
scenarios in~\cite{csdot,bim}). When the second term dominates $\Phi$ and
$\zeta$ become quite different,
such as in cyclic scenarios, and here. Inserting (\ref{zeta2norder})
in (\ref{zetaprime})
implies that
\be
\Phi\propto k^{-5/2} {\cal H}
\ee
so that the spectral index for $\Phi$ is $n_s=-1$ (a red spectrum) and 
it diverges like ${\cal H}\sim \sqrt{\ep_0}$ as we approach the
crunch.

The situation is similar, but slightly better, 
to that of a collapsing matter Universe~\cite{wands}.
In that case fluctuations in $\zeta$ are scale-invariant, but those in 
$\Phi$ are red (with $n_s=-3$). However in such models severe fine
tuning afflicts the amplitude, because the fluctuations in $\zeta$ 
and $\Phi$ diverge as we approach the crunch. That problem doesn't
plague out model, at least if we adopt the view that it's the fluctuations
in $\zeta$ that matter. Still, the potential behaves somewhat pathologically.

Needless to say that this model shares with all other cyclic
scenarios the usual
uncertainties about the transmission of the spectrum past the bounce.

\section{Conclusions}
We considered fluctuations in adiabatic hydrodynamical 
models with varying $w$. Our results can be summarised as follows. 

Under the adiabatic assumption, inflation can never be realized 
(since it implies $w>0$ for any sustained 
period). But neither can the simplest realizations of the 
varying speed of sound mechanism: power-law equations of state  
in expanding universes.
For adiabatic fluids $c_s^2$ and $w$ must then be proportional. 
The variation in $w$ has effects on the varying mass term $z''/z$
describing the effects of expansion. The effect of $c_s$ on the
pressure cancels with the effect of $w$ on $z''/z$ and so the 
horizon problem is in fact never solved. 

To bypass this result in 
expanding Universes one must drop the assumption that $w$ is 
a power-law. A sharp phase transition in $c_s$ leads to scale-invariance
if the initial conditions are set by a thermal state. A model with $w\ll 1$
and vacuum fluctuations is also a possibility, but the scales produced
must then be stretched by an external mechanism. 

For contracting universes the vistas expand. 
Specifically, in cyclic and ekpyrotic models, we find that the law
$c_s^2=3w\propto \rho^2$ leads to scale invariant fluctuations. 
Some of the problems in the ekpyrotic scenario are even bypassed.
Scale-invariance is achieved in variable $\zeta$ and is present in both
of its modes. The amplitude of the fluctuations in $\zeta$ 
freezes-in, instead of diverging. However the spectrum of fluctuations in
$\Phi$ is red and its amplitude time-dependent. 
The problem of the transmission of these
spectra to life after the bounce remains open and unsolved. 

One may be rightly concerned that causality paradoxes afflict these
models, given that $c_s>1$ is permitted, a matter that affects any varying
speed of sound scenario. As discussed in~\cite{bim} a resolution of
these problems is achieved by a bimetric reformulation. 
The larger speed of sound in the gravity frame then
simply signals the presence of a non-conformal matter metric. 
In previous work (based on the anti-DBI action~\cite{bim}) the
re-examination of the model in terms of two metrics only reinforced
its motivation, as it resulted from the minimal bimetric theory.
Examining the bimetric structure behind the models discussed here
is therefore relevant, but beyond the scope of this paper. In a 
future publication we shall examine the status of perfect fluids
in bimetric theories where the bi-scalar is a spectator 
field~\cite{futurebim}.
In another forthcoming paper~\cite{nonad} we shall also 
examine the implications of dropping the
assumption of adiabaticity pervading this paper.

{\bf Acknowledgments} We'd like to thank an anonymous referee for very
helpful comments and for correcting several typos and STFC for financial 
support.


\begin{thebibliography}{99}
\bibitem{hu}W. Hu, Astrophys.J. 506: 485, 1998. 
\bibitem{jochen}J. Weller and A. Lewis, M.N.R.A.S. 346: 987, 2003. 
\bibitem{infl}A. Guth, {\it Phys.Rev.} {\bf D23} 347 (1981); 
A. Linde, {\it Phys. Lett} {\bf B 108}, 1220 (1982). 
\bibitem{ekp}P. Steinhardt and N. Turok, Science 296: 1436-1439, 2002.
\bibitem{hag}A. Nayeri, R. Brandenberger and C. Vafa, Phys. Rev.
Lett.97: 021302, 2006.
\bibitem{csdot}J. Magueijo, Phys. Rev. Lett. 100, 231302 (2008).
\bibitem{piazza}J. Khoury and F. Piazza, arXiv:0811.3633.
\bibitem{chap}N. Billic, G. Tupper and R. Viollier, Phys.Lett. B535,
17 ,2002; 
M. Bento, O. Bertolami and A. Sen, Phys.Rev. D66: 043507, 2002. 
\bibitem{barr}J. Barrow, Phys. Lett B235, 1990.
\bibitem{and}J. Barrow and A. Liddle, Phys.Rev. D47, 1993;
Phys.Rev.D74:127305,2006. 
\bibitem{staro}A. Starobinsky, JETP Lett. 82,2005.
\bibitem{pedro}P. Ferreira and J. Magueijo, 
Phys.Rev. D78: 061301,2008. 
\bibitem{nonad}J. Noller and J. Magueijo, 0911.1907.
\bibitem{futurebim}J. Noller and J. Magueijo, in preparation.
\bibitem{durrer}R. Durrer, Phys. Rev. D 42, 2533 (1990).
\bibitem{stew}J. M. Stewart, Class.Quant.Grav. 7: 1169-1180, 1990. 
\bibitem{bardeen} J. Bardeen, Phys. Rev. D 22,
1882 (1980).
\bibitem{kodama}
H. Kodama and M. Sasaki, Prog. Theor. Phys. Suppl.78: 1-166, 1984. 
\bibitem{lidsey}J. Lidsey et al, Rev. Mod. Phys. 69: 373-410, 1997.
\bibitem{mukh} S. Mukhanov, H. Feldman and R. Brandenberger, Phys.Rept. 215: 203-333, 1992; {\it Physical Foundations of Cosmology}, V. Mukhanov, CUP (2005).
\bibitem{amend}C. Armendariz-Picon and E. Lim, JCAP 0312: 002, 2003;
C. Armendariz-Picon JCAP 0610: 010, 2006.
\bibitem{am}A. Albrecht and J. Magueijo,
{\it Phys.Rev.} {\bf D 59} 043516 (1999).
\bibitem{holo}J. Magueijo, L. Smolin and C. Contaldi, Class.Quant.Grav.24:
3691-3700, 2007.
\bibitem{bim}J. Magueijo, Phys.Rev.D79: 043525, 2009.
\bibitem{ekpperts}S. Gratton et al, Phys.Rev. D69 (2004) 103505.
\bibitem{multiekp}
E.~I.~Buchbinder, J.~Khoury and B.~A.~Ovrut,
Phys.\ Rev.\  D {\bf 76}, 123503 (2007);
J. Lehners et al. Phys. Rev. D 76, 103501 (2007);
P. Creminelli and L. Senatore, JCAP 0711, 010 (2007);
K. Koyama and D. Wands, JCAP 0704, 008 (2007); 
K. Koyama, S. Mizuno and D. Wands, Class. Quant. Grav. 24, 3919 (2007).
\bibitem{wands}F. Finelli and R. Brandenberger, Phys. Rev. D 65, 103522 
(2002); D. Wands, Phys. Rev. D 60, 023507 (1999).

\end{thebibliography}
\end{document}